\documentclass[preprint,5p,twocolumn,pdflatex]{elsarticle}

\usepackage{amssymb}
\usepackage{amsmath}

\newcounter{bla}

\journal{Computer Physics Communications}

\begin{document}

\begin{frontmatter}



\title{FermiSurfer: Fermi-surface viewer providing multiple representation schemes}


\author[a]{Mitsuaki Kawamura\corref{author}}

\cortext[author] {Corresponding author.\\\textit{E-mail address:} mkawamura@issp.u-tokyo.ac.jp}
\address[a]{Institute for Solid State Physics, 
  The University of Tokyo, Kashiwa 277-8581, Japan}

\begin{abstract}
  FermiSurfer is a newly developed Fermi-surface viewer designed
  to facilitate the understanding of the physical properties of metals.
  It can display the Fermi surfaces of a material, color plots of arbitrary $k$-dependent quantities,
  the Fermi surface at arbitrary cross sections (Fermi lines),
  cross- or parallel- eye three-dimensional stereogram views,
  nodal lines, extremal orbits, and highlight the occupied or empty side of the Fermi surface.
  In addition, various first-principles software packages can produce input for FermiSurfer.
  This paper explains how to use FermiSurfer and demonstrates its usefulness by investigating
  the origin of the anisotropic superconductivity of YNi$_2$B$_2$C.
\end{abstract}


\end{frontmatter}



{\bf PROGRAM SUMMARY}

\begin{small}
\noindent
{\em Program Title:} FermiSurfer \\
{\em Licensing provisions:} MIT \\
{\em Programming language:} C \\
{\em Nature of problem:}
The anisotropy of quantities on Fermi surfaces (the character of orbitals, etc.) strongly affects
electronic properties, such as superconductivity or
thermoelectricity, in metals.
Observing the anisotropy, however, remains difficult.
Thus, a graphical tool that can elucidate how quantities vary
over complicated Fermi surfaces is highly desirable.
\\
{\em Solution method:}
FermiSurfer \cite{fermisurfer} uses the tetrahedron method \cite{doi1991efficient} to compute Fermi surfaces.
The French-curve interpolation \cite{Akima:1970:NMI:321607.321609} is used to display smooth Fermi surfaces. 
\\
{\em Additional comments including Restrictions and Unusual features :}  
The parallel- and cross- eye stereograms are available to enhance the visibility
of complicated Fermi surfaces.
This program works on any operating system, such as Linux, UNIX, macOS, and Windows,
in which OpenGL library \cite{opengl} is installed.
\\
\\
\end{small}

\section{Introduction} \label{sec_intro}
The Fermi surface is called ``the face of metal'' \cite{springford2011electrons}
and strongly affects the properties of metals
because it is the most active region in reciprocal space.
The shape of the Fermi surface affects the oscillation of
physical quantities that lead, for example, to
the de Haas--van Alphen effect and the Shubnikov--de Haas effect
\cite{kittel2004introduction}.
Almost-parallel Fermi surfaces cause anomalies in the response functions
at the corresponding wave vector (i.e., the nesting vector) \cite{0305-4608-3-4-022}.
Alkali metals have a spherical Fermi surface,
and the electrons in these metals behave as a nearly free electrons.
Conversely, the Fermi surfaces of cuprates are almost two dimensional
and some of them are strongly nested \cite{RevModPhys.61.433}.
MgB$_2$ \cite{ISI:000167194300040} has both two- and three-dimensional
Fermi surfaces, with significantly different superconducting gaps at each surface
\cite{PhysRevLett.89.187002,ISI:000177428000033,PhysRevB.66.020513}.

By using software such as XCrysDen \cite{KOKALJ2003155},
Fermi surfaces may be displayed as unicolor isosurfaces to reflect the single-electron energy and
we can find the nesting vector.
However,  because these plots show only the shape of the Fermi surface,
their information content is limited.
To obtain more information requires the ability to
plot arbitrary $k$-dependent quantities on Fermi surfaces,
where $k$ is a Bloch wave vector.
Although VESTA \cite{Momma:db5098}, which is a program for visualizing
the crystalline structure and charge density of materials, can draw such plots,
a VESTA Fermi-surface plot especially for multiband systems involves a complicated procedure.
Also, VESTA cannot display the first Brillouin zone.

To address this situation, we developed a program called ``FermiSurfer'' \cite{fermisurfer}
to display arbitrary $k$-dependent quantities on the Fermi surface of a given material. 
In addition, FermiSurfer can plot the cross-section of Fermi surfaces,
stereograms, extremal orbits and nodal lines of $k$-dependent quantities.
The program reads a simple input file that can be generated by arbitrary  software and
is distributed through the MIT X consortium license,
so it can be used and included in all software such as Winmostar \cite{winmostar}.
For example, the first-principles software packages, 
{\sc Quantum ESPRESSO} \cite{0953-8984-29-46-465901}
(the plane-wave- and pseudopotential-based code)
and
Superconducting-Toolkit \cite{PhysRevB.95.054506}
based on the density functional theory for superconductors \cite{PhysRevB.72.024545},
can both generate files readable by FermiSurfer.
This paper provides an introduction to FermiSurfer.
To begin, Sec. \ref{sec_method} explains the method used by the program, and then 
Sec. \ref{sec_install} shows how to install it.
Section \ref{sec_input} discusses the input-file format,
Sec. \ref{sec_running} shows how to use its functions, and
Section \ref{sec_example} shows an example of how FermiSurfer may be used in a study
and, finally, Sec. \ref{sec_summary} concludes and summarizes.

\section{Method} \label{sec_method}

In this section we explain the tetrahedron method to compute the fragment of the Fermi surfaces
from the grid data of the orbital energy $\varepsilon_k$.
We also explain the interpolation method to smooth the displayed Fermi surfaces.
In this section we suppress the band index.

\subsection{Tetrahedron method applied to patches}

To draw a Fermi surface with the tetrahedron method \cite{doi1991efficient},
FermiSurfer first reads the orbital energy $\varepsilon_k$ and
any $k$-dependent quantity $X_k$ on a uniform $k$ grid.
The Brillouin zone is then separated into cells, with each cell cut into six tetrahedra
[Fig. \ref{fig_tetra}(a)], and
$\varepsilon_k$ and $X_k$ are linearly interpolated to analytically compute 
triangular isosurface fragments.
Next, in each tetrahedron, we cut out one or two triangles where
$\varepsilon_{k} = \varepsilon_{\rm F}$
and evaluate $X_{i}^{\rm tri}$ ($i=1,2,3$) at the corners of each triangle:
\begin{align}
  X_{i}^{\rm tri} = \sum_{j=1}^4 F_{i j}(
  \varepsilon_{1}, \cdots, \varepsilon_{4}, \varepsilon_{\rm F}) 
  X_{j},
\end{align}
where $\varepsilon_{1}, \cdots, \varepsilon_{4}$ and $X_{1}, \cdots, X_{4}$
are the orbital energy and $k$-dependent quantity
at each corner of the tetrahedron, respectively.
$F_{i j}$ is calculated as follows 
[$a_{i j} \equiv (\varepsilon_i - \varepsilon_j)/(\varepsilon_{\rm F} - \varepsilon_j)$.
we assume $\varepsilon_1 \leq \varepsilon_2 \leq \varepsilon_3 \leq\varepsilon_4$]:
\begin{figure}[!tb]
  \begin{center}
    \includegraphics[width=8cm]{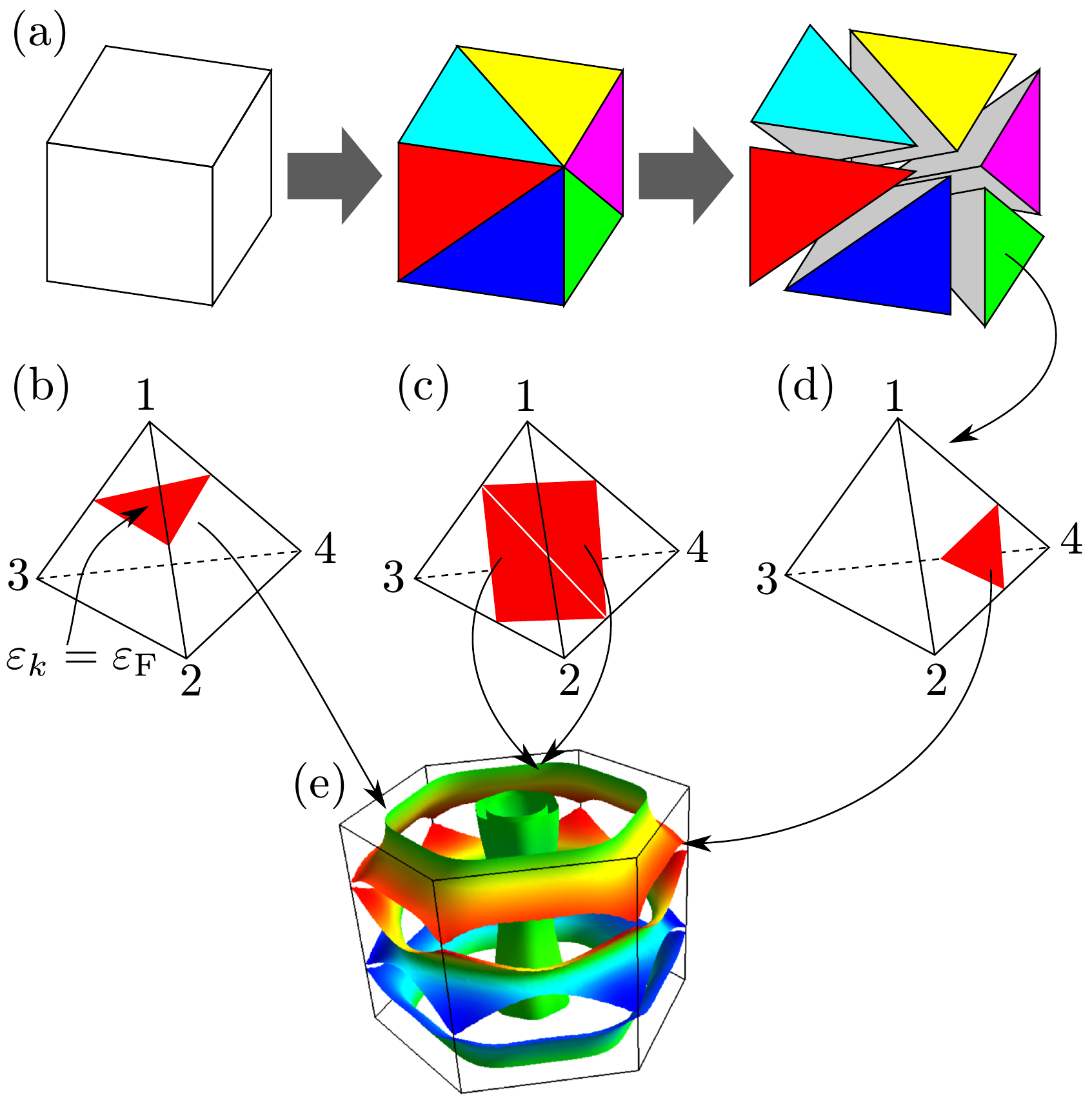}
    \caption{\label{fig_tetra}
      (a) A cell of uniform $k$-point grid is cut into six tetrahedra.
      Dividing a tetrahedron for
      (b) $\varepsilon_1 \leq \varepsilon_{\rm F} \leq \varepsilon_2$,
      (c) $\varepsilon_2 \leq \varepsilon_{\rm F} \leq \varepsilon_3$,
      (d) $\varepsilon_3 \leq \varepsilon_{\rm F} \leq \varepsilon_4$.
      (e) Triangular patches are combined into Fermi surfaces.
    }
  \end{center}
\end{figure}

\begin{itemize}

\item When $\varepsilon_1 \leq \varepsilon_{\rm F} \leq \varepsilon_2 \leq \varepsilon_3 \leq\varepsilon_4$
  [Fig. \ref{fig_tetra}(b)], 

  \begin{align}
    F &= 
    \begin{pmatrix}
      a_{1 2} & a_{2 1} &       0 & 0 \\
      a_{1 3} &       0 & a_{3 1} & 0 \\
      a_{1 4} &       0 &       0 & a_{4 1}
    \end{pmatrix}
  \end{align}
  
\item When $\varepsilon_1 \leq \varepsilon_2 \leq \varepsilon_{\rm F} \leq \varepsilon_3 \leq\varepsilon_4$
  [Fig. \ref{fig_tetra}(c)], 
  \begin{align}
    F &= 
    \begin{pmatrix}
      a_{1 3} &       0 & a_{3 1} & 0 \\
      a_{1 4} &       0 &       0 & a_{4 1} \\
            0 & a_{2 4} &       0 & a_{4 2}
    \end{pmatrix}
  \end{align}
  
  \begin{align}
    F &= 
    \begin{pmatrix}
      a_{1 3} &       0 & a_{3 1} & 0 \\
            0 & a_{2 3} & a_{3 2} & 0 \\
            0 & a_{2 4} &       0 & a_{4 2}
    \end{pmatrix}
  \end{align}

\item When
  $\varepsilon_1 \leq \varepsilon_2 \leq \varepsilon_3 \leq \varepsilon_{\rm F} \leq \varepsilon_4$
  [Fig. \ref{fig_tetra}(d)], 

  \begin{align}
    F &= 
    \begin{pmatrix}
      a_{1 4} &       0 &       0 & a_{4 1} \\
      a_{1 3} & a_{2 4} &       0 & a_{4 2} \\
      a_{1 2} &       0 & a_{3 4} & a_{4 3}
    \end{pmatrix}
  \end{align}

\end{itemize}

We compute the $k$ points and $X_k$ at the corner of each triangular patch as shown above
then paint each patch with colors linearly interpolated from those at the corner.
The normal vector for simulating the lighting conditions is obtained from the Fermi velocity
computed by using the central-difference method.
Finally these triangular patches are combined into Fermi surfaces [Fig. \ref{fig_tetra}(e)].

\label{sec_french}
\subsection{French-curve interpolation}

To smooth the Fermi surfaces for display, we interpolate the energy $\varepsilon_k$ and
the $k$-dependent quantity $X_{nk}$ into a $k$ grid  that is denser than that of the input.
Because of band crossing, the spline,
Fourier, and polynomial interpolations are not appropriate for this purpose;
oscillation occurs even at points far from the band-crossing point
where the energy band has a kink.
Although the disentangled Wannier interpolation \cite{PhysRevB.65.035109} is one of the
most accurate methods to interpolate band structure,
this method requires additional information (i.e. Bloch functions) and
care with the parameters (energy window and projectors), 
so FermiSurfer has difficultly doing this interpolation automatically.
Therefore, we use a simple French-curve interpolation \cite{Akima:1970:NMI:321607.321609},
which requires only the original function itself and is robust against kinks
in the band structures
because of its compact formula.
By using the French-curve interpolation, we obtain the interpolated function ${\tilde f}(x)$
in $x=[0,1]$ from the original function at four points
[$f(-1)$, $f(0)$, $f(1)$, and $f(2)$] as follows:
\begin{align}
  \label{eq_french1d}
  {\tilde f}(x) = a_{-1}(x) f(-1) + a_0(x) f(0) + a_1(x) f(1) + a_2(x) f(2),
\end{align}
where
\begin{align}
  a_{-1}(x) &= - \frac{ x (1-x)^2}{2},
  \nonumber \\
  a_0(x) &= \left\{ (1-x)^2 + 3 x (1-x) + \frac{x^2}{2} \right\} (1-x),
  \nonumber \\
  a_1(x) &=  \left\{ x^2 + 3(1-x)x + \frac{(1-x)^2}{2} \right\} x,
  \nonumber \\
  a_2(x) &= -\frac{x^2(1-x)}{2}.
\end{align}
This interpolated function ${\tilde f}(x)$ satisfies ${\tilde f}(0)=f(0)$, ${\tilde f}(1)=f(1)$,
${\tilde f}'(0)=\{f(1)-f(-1)\}/2$ and ${\tilde f}'(1)=\{f(2)-f(0)\}/2$
and depends only on these four points.
Therefore, the effect of the kink structure at the band-crossing point
is confined to the vicinity of that point.
We can easily extend Eq. (\ref{eq_french1d}) into the three dimensional case as follows:
\begin{align}
  {\tilde f}(x,y,z) = \sum_{l=-1}^2 \sum_{m=-1}^2 \sum_{n=-1}^2
  a_l(x) a_m(y) a_n(z) f(l,m,n).
\end{align}

\section{Installation} \label{sec_install}

This section explains the procedure for installing FermiSurfer.

\subsection{Executable file for Windows}

The FermiSurfer package contains the source code, sample input files, documents, and
the binary file \verb|bin/fermisurfer.exe| for Windows.
Therefore, it does not need to be built for Windows.
The necessary dynamic link library files are also included.

\subsection{Build by hand}

For UNIX, Linux, and macOS, FermiSurfer has to be built manually.
Since FermiSurfer uses the OpenGL \cite{opengl} library and the OpenGL Utility Toolkit (GLUT),
these libraries first have to be installed as follows:
\\
For Debian and Ubuntu
\begin{verbatim}
$ sudo apt-get install freeglut3-dev
\end{verbatim}
For Red Hat Enterprise Linux and CentOS
\begin{verbatim}
$ sudo yum install freeglut-devel.x86_64
\end{verbatim}
For macOS, these libraries can be installed as part of the Xcode utility.

FermiSurfer can be built and installed by using 
\begin{verbatim}
$ ./configure
$ make
$ make install
\end{verbatim}
The binary files \verb|src/fermisurfer| and \verb|src/bxsf2frmsf|
are then generated and copied into the directory \verb|/usr/local/bin/|.

\section{Input file} \label{sec_input}

This section introduces the format of the FermiSurfer input file and
gives examples of source code in Fortran and C to generate such a file.
Also, we introduce the interface between FermiSurfer and other programs.

\subsection{Input-file format}

\begin{figure}[!b]
  \begin{center}
    \includegraphics[width=7cm]{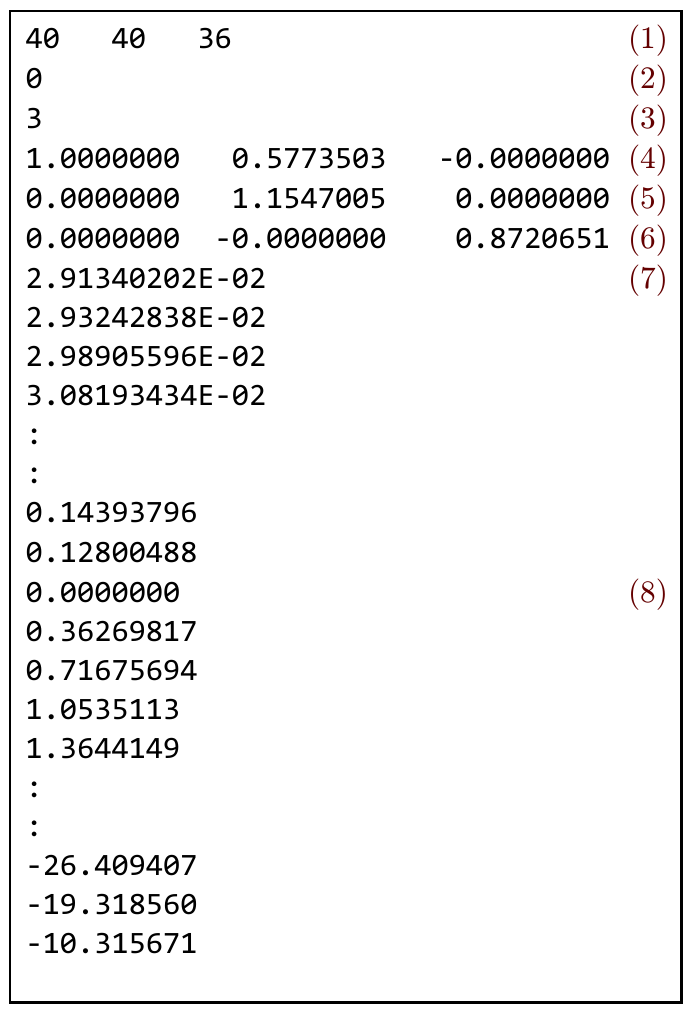}
    \caption{(Color online)
      Example of input file.
      The content labeled by line numbers in parentheses is explained in the main text.
      These labels do not appear in the actual input file.
    }
    \label{fig_input}
  \end{center}
\end{figure}
To compute the Fermi surfaces with a color plot of a $k$-dependent quantity,
FermiSurfer requires the number of $k$ points in three directions,
the reciprocal-lattice vectors, the number of bands,
the orbital energy $\varepsilon_{n k}$ where $n$ is the band index, and
any $k$-dependent quantity $X_{n k}$ (e.g. the superconducting gap).
The quantities $\varepsilon_{n k}$ and $X_{n k}$ are input as grid data.

Figure \ref{fig_input} shows an example input file called \verb|mgb2_vfz.frmsf|
that is contained in the FermiSurfer package.
Each component of this file may be described as follows:
\begin{enumerate}[(1)]
\item The number of $k$-grid points along each reciprocal-lattice vector;
\item Specify the type of $k$ grid (available values are 0, 1, or 2).
  The $k$ grid is represented as follows:
  \begin{align}
    {\boldsymbol k}_{i,j,l} =
    x_{1 i} {\boldsymbol b}_1 + x_{2 j} {\boldsymbol b}_2 + x_{3 l} {\boldsymbol b}_3,
  \end{align}
  where
  $i = 1, \cdots N_1$, $j = 1, \cdots N_2$,  $l = 1 \cdots N_3$,
  and
  $N_1, N_2, N_3$ are the number of $k$ in the direction of each reciprocal-lattice vector.
  Each switch corresponds to the following $x_{\alpha i}$:
  \begin{itemize}
  \item 0: $x_{\alpha i} = (2 i - 1 - N_\alpha)/(2 N_\alpha)$ (i.e., the Monkhorst-Pack grid)
  \item 1: $x_{\alpha i} = (i - 1)/N_\alpha$
  \item 2: $x_{\alpha i} = (2 i - 1)/(2 N_\alpha)$
  \end{itemize}
  
\item The number of bands
\item Reciprocal lattice vector 1
\item Reciprocal lattice vector 2
\item Reciprocal lattice vector 3
\item The orbital energy $\varepsilon_{n k}$.
  By default, FermiSurfer assumes that the Fermi energy is zero.
  We can shift the Fermi energy by using the menu
  ``Shift Fermi Energy'', which is described in Sec. \ref{subsec_menu}.
\item $k$-dependent quantity $X_{n k}$
\end{enumerate}

\begin{figure}[!b]
  \begin{center}
    \includegraphics[width=8cm]{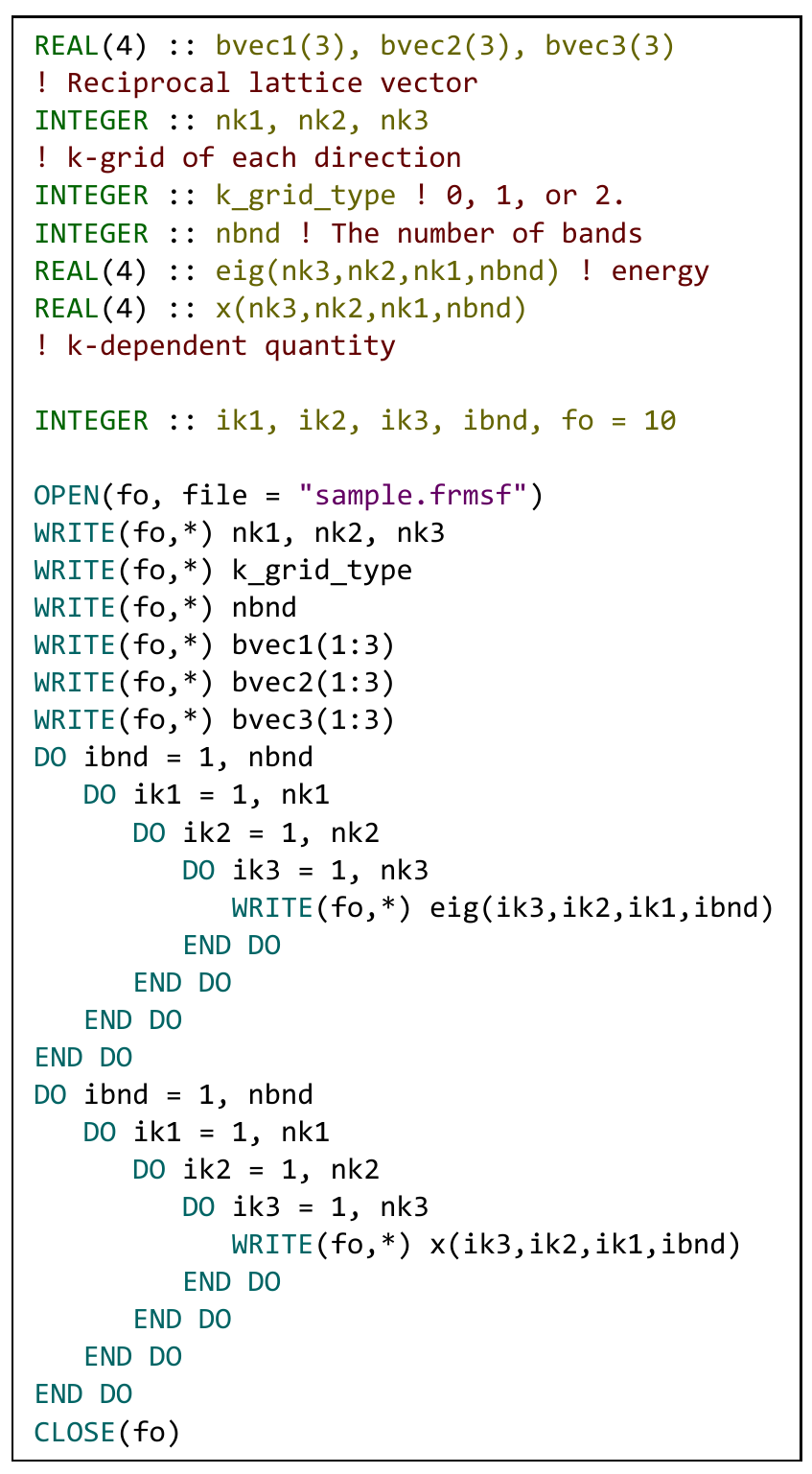}
    \caption{(Color online)
      Example of the Fortran source code to generate input file.
    }
    \label{fig_fortran}
  \end{center}
\end{figure}

\begin{figure}[!b]
  \begin{center}
    \includegraphics[width=8cm]{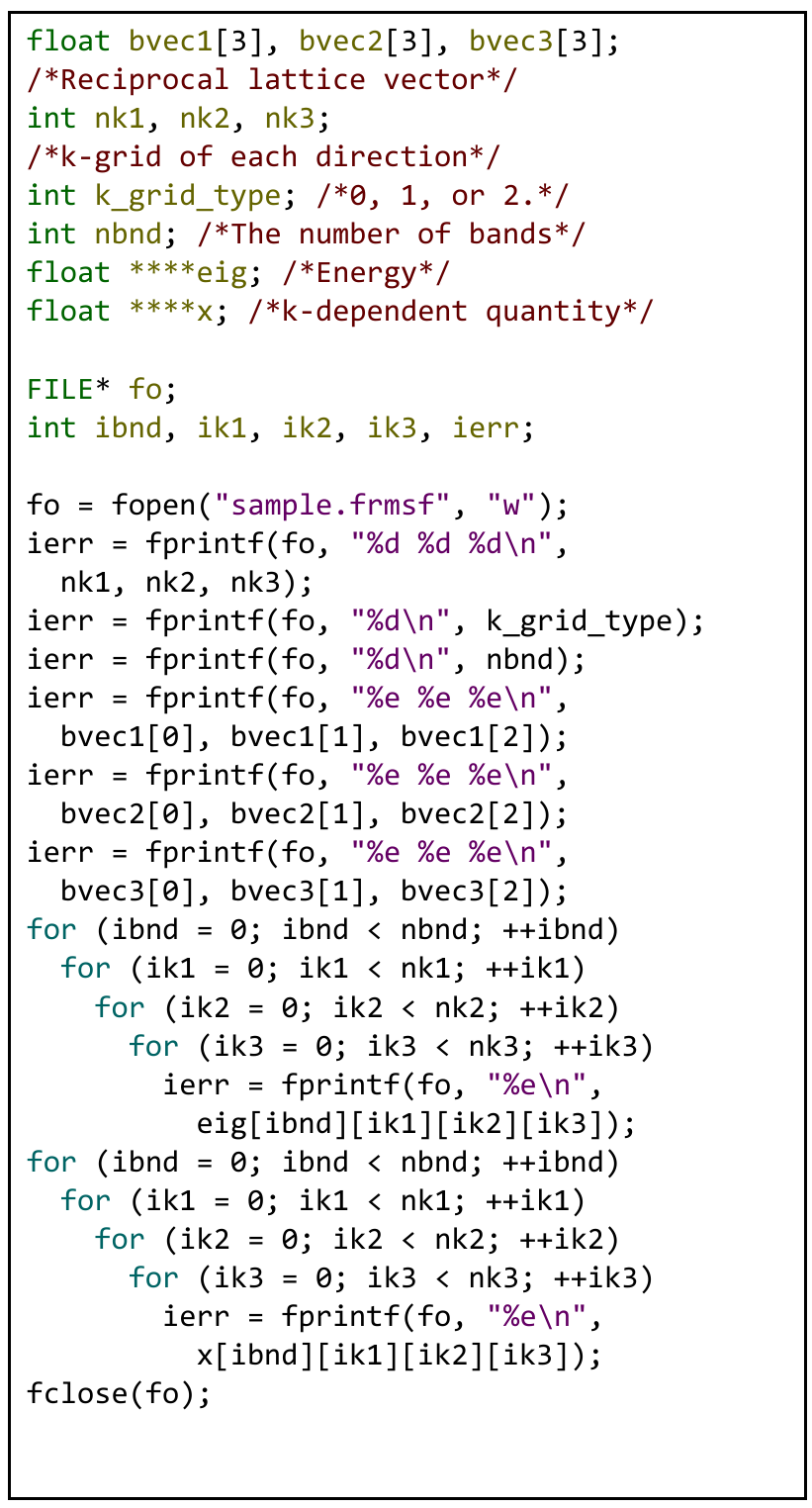}
    \caption{(Color online)
      Example of C source code to generate input file.
    }
    \label{fig_clang}
  \end{center}
\end{figure}

Figures \ref{fig_fortran} and \ref{fig_clang} show the source code
in Fortran and the C, respectively, which generate this input file.

\subsection{Converting BXSF files}

Input files for FermiSurfer can be generated from a XCrysDen 
input file \cite{KOKALJ2003155} (i.e., from the bxsf format)
by using the utility program ``bxsf2frmsf'',
which is included in the FermiSurfer package. 

\subsection{Generating input file with other programs}

There are two first-principles program packages,
{\sc Quantum ESPRESSO} \cite{0953-8984-29-46-465901}
and
Superconducting-Toolkit (SCTK) \cite{PhysRevB.95.054506},
that can output files compatible with FermiSurfer.
{\sc Quantum ESPRESSO} is a first-principles software package which bases its 
calculations on plane waves and pseudopotentials.
Since version 6.2,
this package can generate files containing
the Fermi velocity and the atomic orbital character on Fermi surfaces.
The Fermi velocity on the Fermi surface affects the conductivity and other 
properties, so knowledge of the anisotropy of this velocity
is important for things such as the thermoelectric materials.
Moreover, when the contributions from each atomic orbital vary on the Fermi surface,
various quantities become anisotropic.

SCTK is a software package based on
the density functional theory for superconductors (SCDFT) \cite{PhysRevB.72.024545}.
By using SCTK,
we can compute the superconducting gap $\Delta_{n k}$ and the transition temperature,
and generate files for displaying the superconducting gap with FermiSurfer.
This is done as a postprocess after obtaining the gap function
by self-consistently solving the gap equation of SCDFT.
SCTK can also compute the electron-phonon renormalization
and the screened Coulomb potential that can be displayed
as color plots on Fermi surfaces.
In Sec. \ref{sec_example}, we show an example of study which
is performed by using these programs together with FermiSurfer.

\section{Usage and functions} \label{sec_running}

\subsection{Launch}

For Linux, UNIX, or macOS,
the  executable file is launched as follows: 
\begin{verbatim}
  $ fermisurfer mgb2_vfz.frmsf
\end{verbatim}
A space is required between the command and the name of the input file
(the sample input file \verb|mgb2_vfz.frmsf| contains $z$ elements of
the Fermi velocity in MgB$_2$).

For Windows,
we launch FermiSurfer by right-clicking on the input file,
choosing ``Open With ...'', and
selecting the binary file \verb|fermisurfer.exe|.

\subsection{Appearance}

\begin{figure}[!b]
  \begin{center}
    \includegraphics[width=8cm]{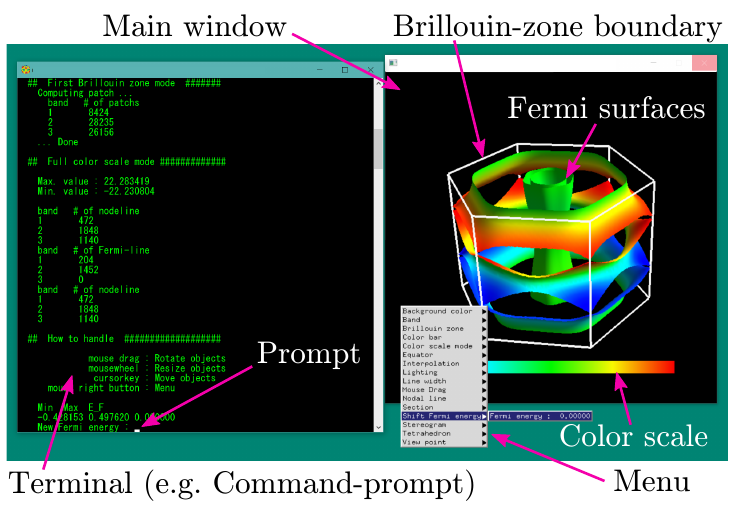}
    \caption{(Color online)
      Main view of FermiSurfer application.
    }
    \label{fig_start}
  \end{center}
\end{figure}
Figure \ref{fig_start} shows the appearance of FermiSurfer during execution.
It uses two windows (the main window and the terminal window):
The main window displays the Fermi surface, the Brillouin-zone boundary,
and the color scale.
A pop-up menu appears upon right-clicking anywhere in the main window.
It is also possible to rotate, resize, and translate the objects by
dragging with the mouse, using the mouse wheel, and using the cursor keys, respectively.
Information about the input data and the figure appear in the terminal window.
Some commands in the pop-up menu, such as ``Shift Fermi energy'', require
input at the prompt in the terminal window.

\subsection{Menu} \label{subsec_menu}

We now explain the commands in the pop-up menu.

\noindent \underline{\bf Background color}

The background (Brillouin zone boundary) color is toggled
between black and white (white and black).

\noindent \underline{\bf Band}

Each band can be turned on and off separately.
The bands displayed are marked with an ``x''
in the subcommand of this command. 

\noindent \underline{\bf Brillouin zone}

\begin{figure}[!b]
  \begin{center}
    \includegraphics[width=8cm]{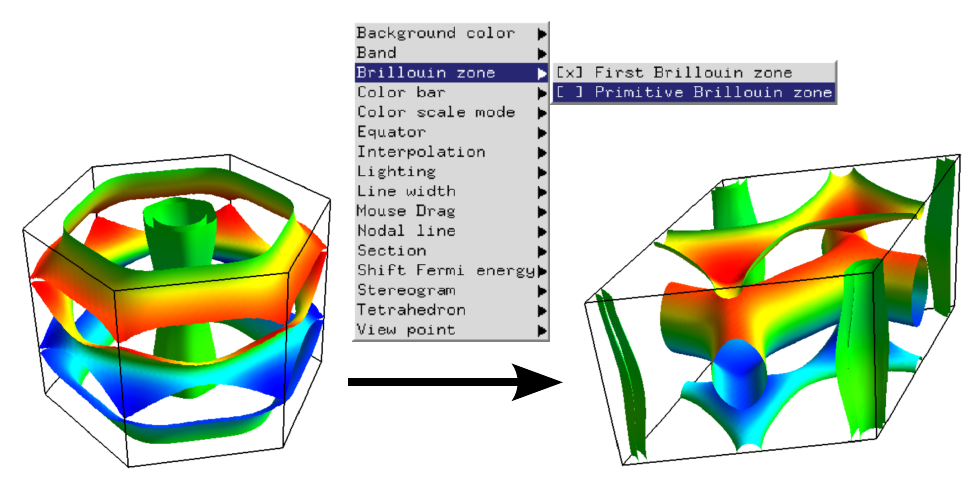}
    \caption{(Color online)
      The type of Brillouin zone is changed by using the ``Brillouin zone'' command. 
      \label{fig_brillouinzone}
    }
  \end{center}
\end{figure}

The type of Brillouin zone is either the first Brillouin zone or the 
primitive Brillouin zone (see Fig. \ref{fig_brillouinzone}).
The first Brillouin zone is the region surrounded by the Bragg planes nearest to the 
${\rm \Gamma}$ point, whereas the primitive Brillouin zone is
a parallelepiped whose edges are the input reciprocal-lattice vectors.

\noindent \underline{\bf Color bar}

This command displays and hides the color scale.

\noindent \underline{\bf Color scale mode}

This command modifies the color scale used to display quantities on the Fermi surface.
It has eight subcommands 
``Auto'', ``Manual'', ``Unicolor'', ``Fermi velocity (Auto)'',
``Fermi velocity (Manual)'', ``Grayscale (Auto)'', and ``Grayscale (Manual)''.
The subcommands ``Auto'' and ``Manual'' allow
the input physical quantity to be plotted in color.
``Gray scale (Auto)'' and ``Gray scale (Manual)'' plot the input physical quantity
in Gray scale.
The ``Unicolor'' subcommand paints the Fermi surface in a single color.
The ``Fermi velocity (Auto)'' and ``Fermi velocity (Manual)'' subcommands
produce color plots of the Fermi velocity computed 
internally from the energy bands. 
Note that ``Unicolor'', ``Fermi velocity (Auto)'', and ``Fermi velocity (Manual)'' 
ignore the input physical quantity. 
``Auto'', ``Fermi velocity (Auto)'', or ``Gray scale (Auto)'' automatically set
the blue and red extremities of the color scale to
the minimum and maximum values of the displayed quantity, respectively.
Conversely, subcommands ``Manual'', ``Fermi velocity (Manual)'', and ``Gray scale (Manual)''
allow the user to set the values for the blue and red extremities of the color scale. 
     
\noindent \underline{\bf Equator}

\begin{figure}[!tb]
  \begin{center}
    \includegraphics[width=9cm]{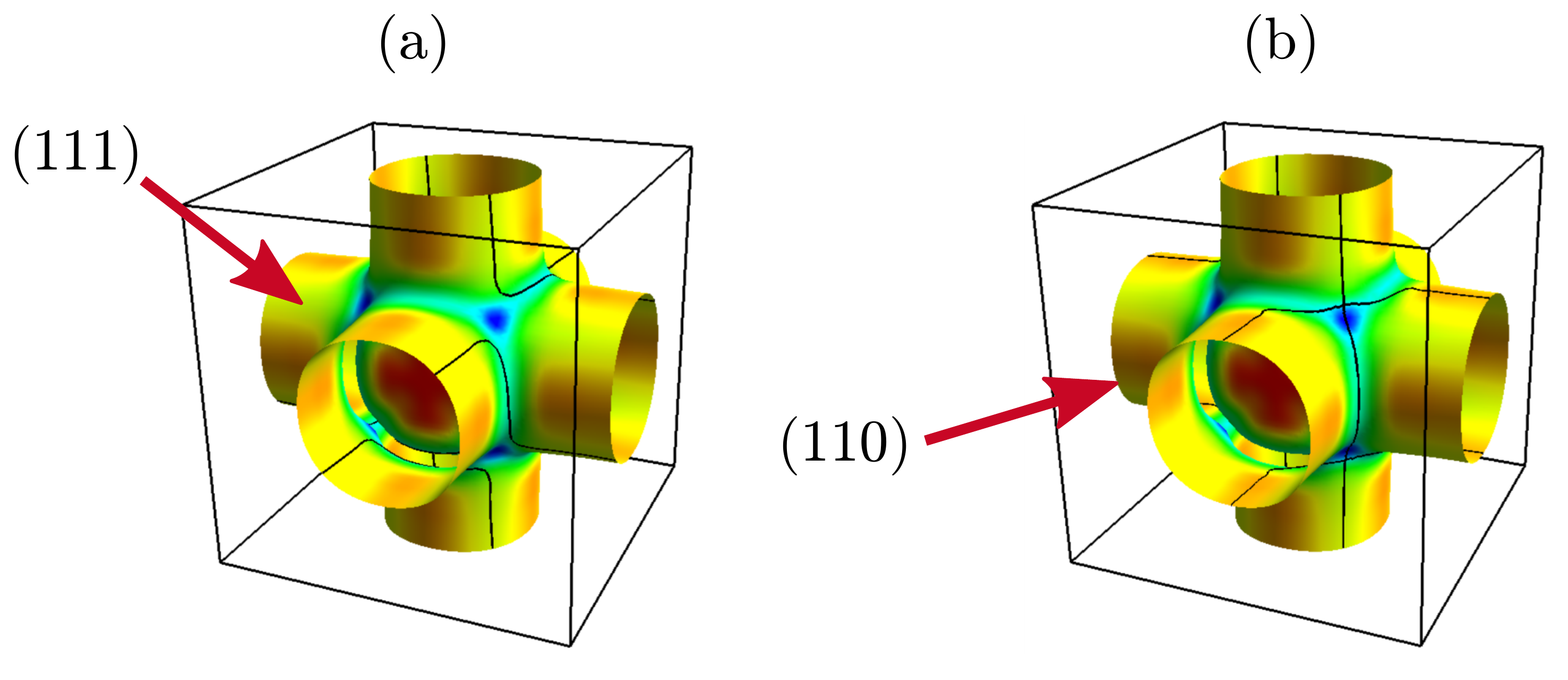}
    \caption{\label{fig_equator}
      Fermi surfaces of SrVO$_3$ together with color plot of Fermi velocity.
      Red, green, and blue indicate the maximum, middle, and minimum
      Fermi velocity, respectively.
      Black lines on the Fermi surface indicate
      the extremal orbits along the (a) (111) direction and
      (b) (110) direction.
    }
  \end{center}
\end{figure}

In experiments such as ultrasonic attenuation \cite{schrieffer1983theory} and
the de Haas--van Alphen oscillation \cite{kittel2004introduction},
the electronic states situated on lines where the Fermi velocity is orthogonal to a vector
(e.g. orientation vector of the magnetic field) dominate the response;
these lines are called the extremal orbits.
By visualizing quantities such as electron-phonon renormalization on the extremal orbits,
we can study the anisotropy appearing in these experiments.
Using the ``Equator'' command, the user can toggle the display of the extremal orbits and modify
the vector to be orthogonalized to the Fermi velocity.
As an example, Fig. \ref{fig_equator} shows the Fermi surfaces of SrVO$_3$
together with the extremal orbits in the (111) and (110) directions
and a color plot of the Fermi velocity.
This graphic shows the (110) orbit passes through the region of minimum Fermi velocity,
whereas the (111) orbit does not pass through that region.

\noindent \underline{\bf Interpolation}

This command smooths the Fermi surface by using the French-curve interpolation (Sec. \ref{sec_french}).
This command modifies the number of extra points that are computed by using the interpolation.

\noindent \underline{\bf Lighting}

\begin{figure}[!tb]
  \begin{center}
    \includegraphics[width=9cm]{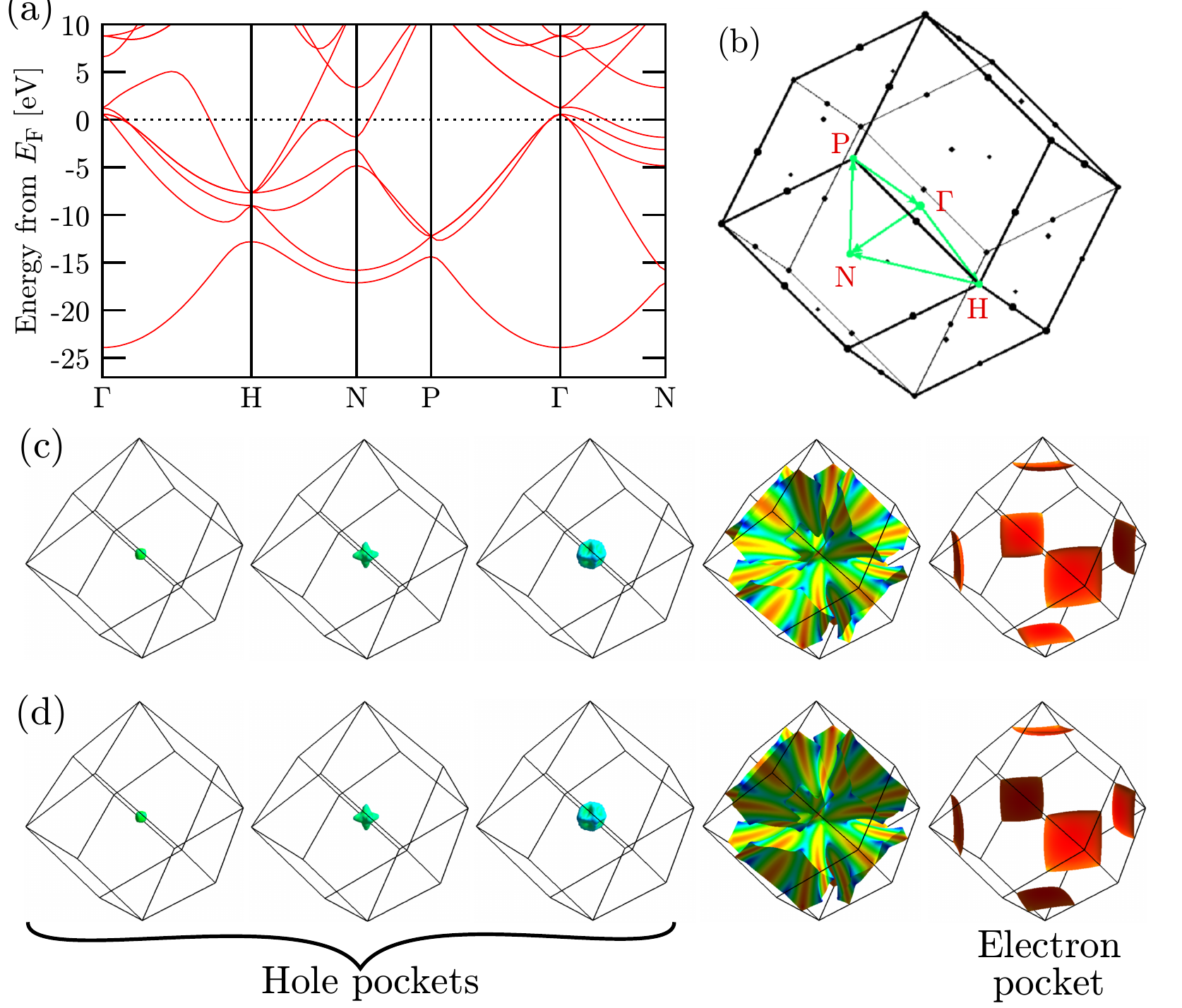}
    \caption{\label{fig_lighting}
      (a) Band structure of H$_3$S at 150 GPa (Im$\bar{3}$m phase) 
      and $k$-point path for this plot (b),
      (c) Fermi surfaces of H$_3$S with both sides illuminate.
      The color scale gives the absolute value of the Fermi velocity.
      (d) Fermi surfaces of H$_3$S with only the occupied side illuminate.
      The color scale gives the absolute value of the Fermi velocity
      (red and blue color indicate the maximum and the minimum Fermi velocity, respectively).
    }
  \end{center}
\end{figure}

The Fermi surface is the interface between the electronically occupied and unoccupied regions.
Via three commands, FermiSurfer allows the user to choose which side of the Fermi surface is illuminated:
Both sides, the unoccupied side only, or the occupied side only.
Figure \ref{fig_lighting} shows the Fermi surface of 
H$_3$S \cite{drozdov2015conventional, duan2014pressure}
(Im$\bar{3}$m phase) at 150 GPa with only the occupied side illuminated.
The results shows the electron pocket at the corner of the Brillouin zone and 
the hole pockets at the center of the Brillouin zone.

\noindent \underline{\bf Line width}

This command modifies the width of the Brillouin-zone boundary, the nodal line, etc.
   
\noindent \underline{\bf Mouse drag}

This command changes the effect of a mouse-left-drag.
The three associated subcommands ``Rotate'', ``Scale'', and ``Translate''
rotate objects along the mouse drag,
resize objects with an upward or downward drag, and
translate objects along the mouse drag,
respectively.

\noindent \underline{\bf Nodal line}

This command toggles on and off 
the line on which the $k$-dependent quantity is zero (i.e., the nodal line).

\noindent \underline{\bf Section}

\begin{figure}[!b]
  \begin{center}
    \includegraphics[width=9cm]{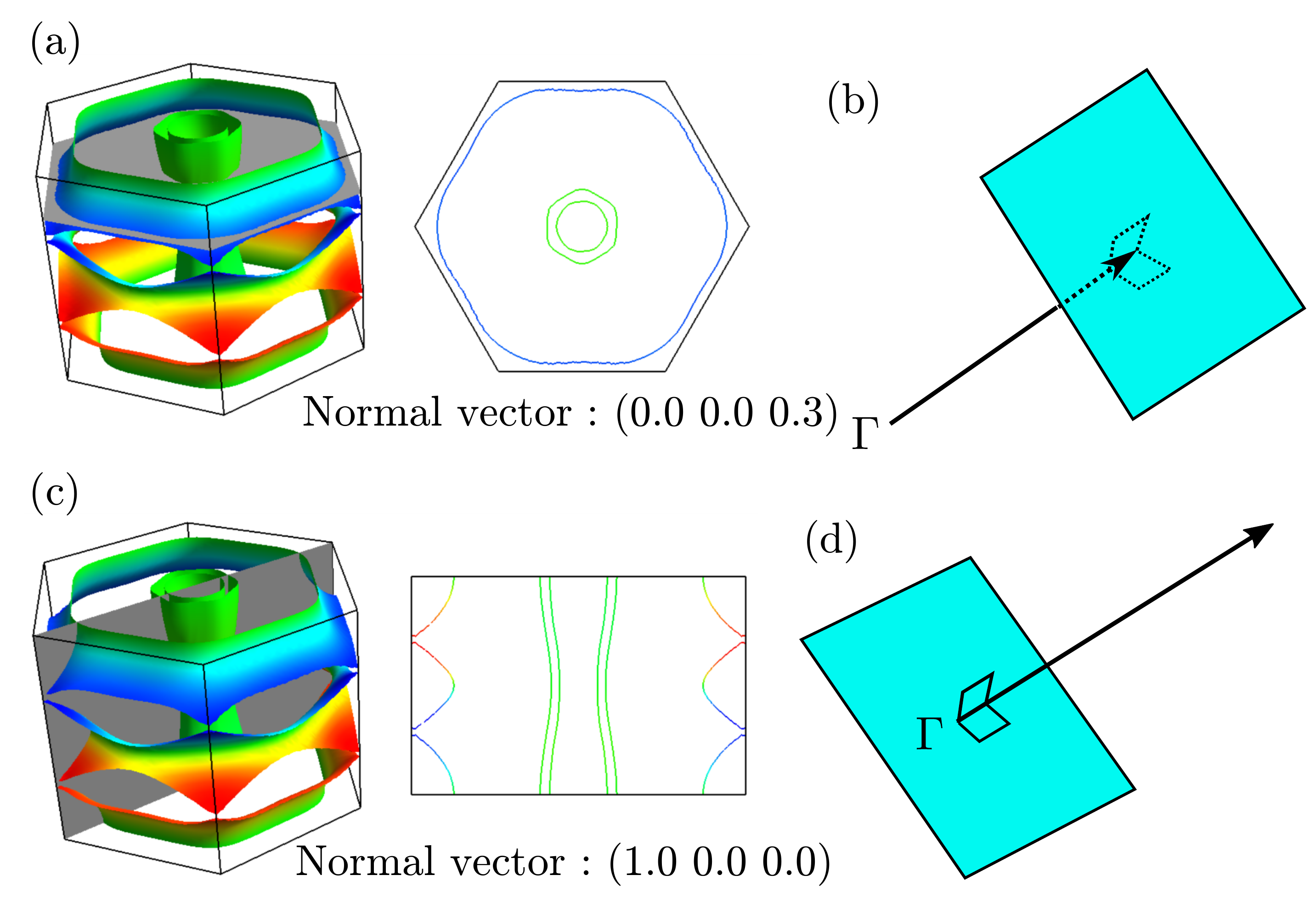}
    \caption{(Color online)
      (a) Fermi surface on cross section perpendicular to vector (0.0, 0.0, 0.3).
      (b) Schematic illustration of cross section used in panel (a). 
      (c) Fermi surface on the cross section containing the $\Gamma$ point
      and perpendicular to vector (1.0, 0.0, 0.0).
      (d) Schematic illustration of cross section used in panel (c).
      \label{fig_section}
    }
  \end{center}
\end{figure}

This command allows the user to display a two-dimensional plot of the Fermi surface
from an arbitrary cross section of the Brillouin zone.
This command has three subcommands: ``Section'', ``Modify section'',
and ``Modify Section (across Gamma)''.
The subcommand ``Section'' toggles on and off the two-dimensional plot of the Fermi surface.
The subcommand ``Modify Section'' and ``Modify Section (across Gamma)''
allow the user to specify the cross section.
With these two subcommands, the user specifies a vector in reciprocal fractional coordinates, and
the cross section is vertical to this vector.
With the ``Modify Section'' subcommand, 
the cross section is perpendicular to the tip of the specified vector [Fig. \ref{fig_section} (a)].
Conversely, 
the ``Modify Section (across Gamma)'' subcommand puts the $\Gamma$ point in 
the cross section [Fig. \ref{fig_section} (c)].
In the latter case, the length of the specified vector is ignored.
     
\noindent \underline{\bf Shift Fermi energy}

This command shifts the Fermi energy (which is zero by default) to an arbitrary value.
This command first displays in the terminal window the current Fermi energy and
the minimum and maximum energy in the input file.
 The user then inputs the new Fermi energy at the prompt
in the terminal window, and the new Fermi surface is displayed.

\noindent \underline{\bf Stereogram}

\begin{figure}[!b]
  \begin{center}
    \includegraphics[width=8cm]{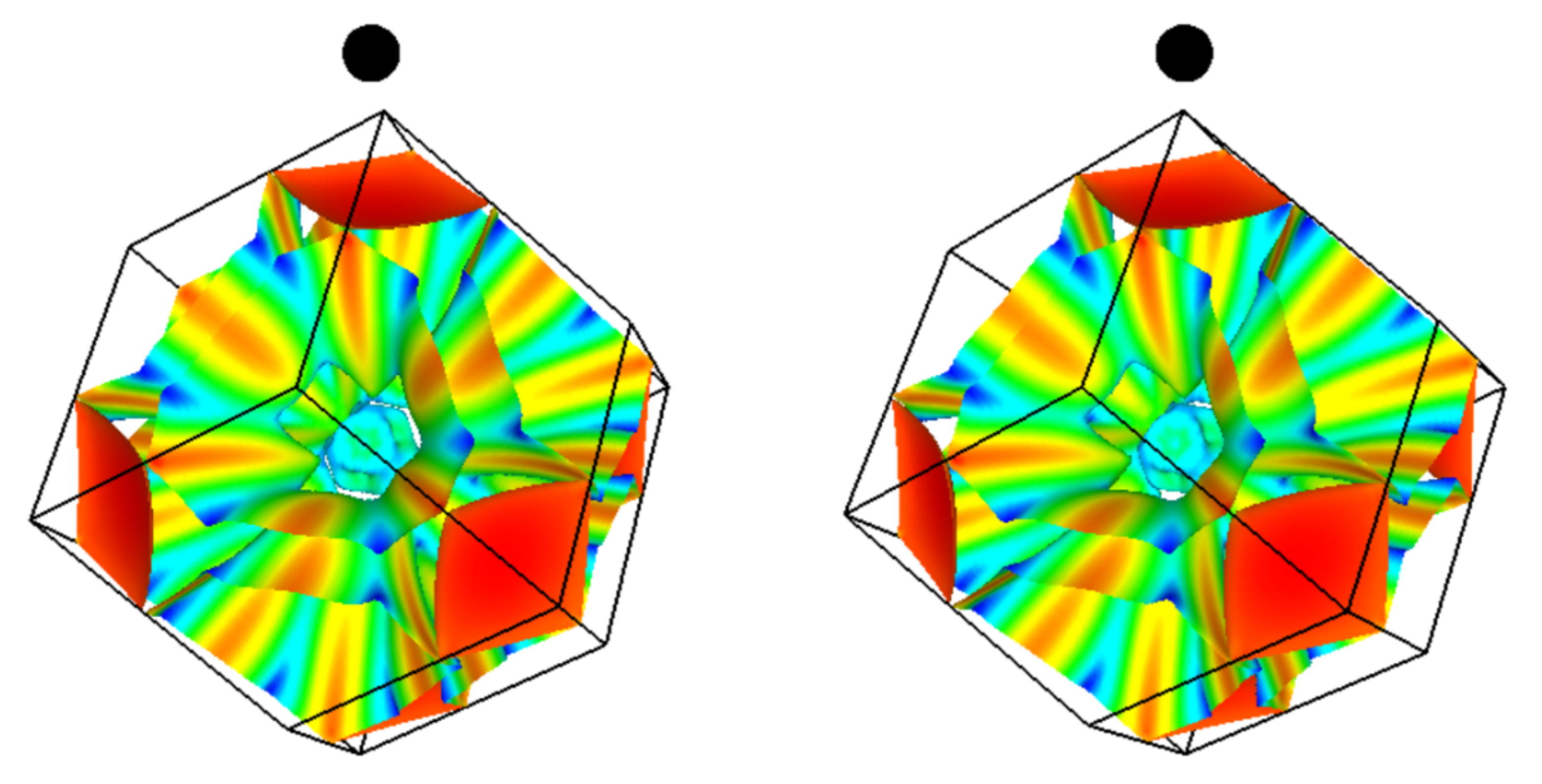}
    \caption{(Color online)
      Parallel-eye stereogram of
      Fermi surface of H$_3$S at 150 GPa (Im$\bar{3}$m phase)
      together with color plot of Fermi velocity.
      \label{fig_stereogram}
    }
  \end{center}
\end{figure}

This command toggles on and off the stereogram
(parallel eye and cross eye; see Fig. \ref{fig_stereogram}).
It contains three subcommands: ``None'', ``Parallel'', and ``Cross''.
In the default setting ``None'', the stereogram is not shown.
The ``Parallel'' and ``Cross'' subcommands display
a parallel- and cross-eye stereogram, respectively.

\noindent \underline{\bf Tetrahedron}

This command allows the user to change the direction in which a 
tetrahedra is cut from a cell [Fig. \ref{fig_tetra} (a)].

\noindent \underline{\bf Viewpoint}

This command changes the viewpoint. It has three subcommands, 
``Scale'', ``Position'', and ``Rotation'', which allow the user to change the size, position, 
and orientation of objects, respectively.
For each subcommand, the current value is first displayed,
then the user is prompted to input a new value in the terminal window. 

\section{Examples} \label{sec_example}

\begin{figure}[!tb]
  \begin{center}
    \includegraphics[width=9cm]{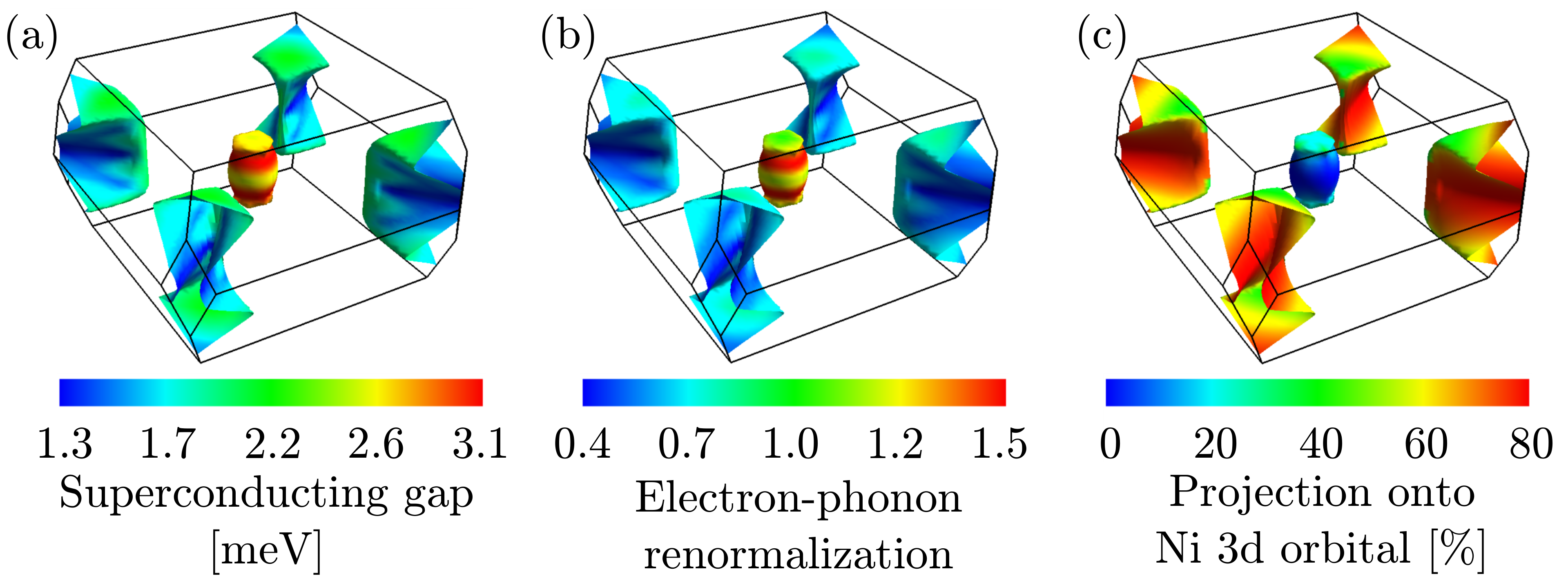}
    \caption{\label{fig_yni2b2c}
      (Color online) Part of Fermi surface of YNi$_2$B$_2$C together with color plots of
      (a) superconducting gap,
      (b) electron-phonon renormalization, and
      (c) Ni 3d character.
    }
  \end{center}
\end{figure}

This section discusses the use of FermiSurfer to investigate the anisotropic superconductivity
in YNi$_2$B$_2$C \cite{Mazumdar1993413,ISI:A1994MR49400048}.
Although this material is a conventional phonon-mediated superconductor,
it has highly anisotropic superconducting gaps
\cite{PhysRevB.67.014526,PhysRevLett.92.147002,PhysRevLett.89.137006,PhysRevB.81.180509}.
Figure \ref{fig_yni2b2c}(a) shows the superconducting gap computed with SCTK
\cite{PhysRevB.95.054506}.
Small-gap (large-gap) regions appear at the corners 
(center) of the Brillouin zone. The same 
tendency appears in the color plot of the
electron-phonon renormalization [Fig. \ref{fig_yni2b2c}(b)]
\begin{align}
  \lambda_{n k} = \sum_{q \nu n'} \frac{2}{\omega_{q \nu}}|g_{n k n' k+q}^{q \nu}|^2
  \delta(\varepsilon_{n' k+q}-\varepsilon_{\rm F}),
\end{align}
where 
$\omega_{q \nu}$ is the phonon frequency for wave number $q$ and branch $\nu$,
$\varepsilon_{n k}$ is the Kohn--Sham eigenvalue for wave number $k$ and band index $n$,
$g_{n k n' k+q}^{q \nu}$ is the electron-phonon vertex between
Kohn--Sham orbitals $(n,k)$, $(n',k+q)$ and the phonon $(\nu, q)$.
Therefore, the anisotropy of the superconducting gaps comes from the anisotropy of
the electron-phonon interaction.
The origin of the anisotropy of the electron-phonon interaction is traced as follows:
Figure \ref{fig_yni2b2c} (c) shows the projection of the Kohn--Sham orbitals onto
the Ni 3d orbitals,
\begin{align}
  \sum_{\tau={\rm Ni}1,{\rm Ni}2} |\langle \varphi_{n k} | \phi_{\tau, n=3, l=2, m} \rangle|^2,
\end{align}
where $\varphi_{n k}$ is the Kohn--Sham orbital for the wave number $k$ and band index $n$,
and $\phi_{\tau n l m}$ is the atomic orbital whose principal,
angular momentum, and azimuthal quantum numbers are $n, l, m$, respectively.
Based on this and the previous figures, we see that
locations of strong Ni 3d correspond to small electron-phonon interaction, and vice versa.
Further study \cite{PhysRevB.95.054506} indicates that
the Ni 3d state, which is localized around Ni atoms strongly screens this
electron-phonon interaction.
In contrast with YNi$_2$B$_2$C,
MgB$_2$ \cite{ISI:000167194300040} exhibits a separated multiband superconductivity
\cite{PhysRevLett.89.187002,ISI:000177428000033,PhysRevB.66.020513},
so continuous multiband superconductivity occurs in this material because of the variation of the
hybridization on the Fermi surface.

\section{Summary} \label{sec_summary}

We present herein FermiSurfer,
which is a Fermi-surface viewer designed to facilitate an intuitive understanding
of electronic states in metals.
FermiSurfer reads the orbital energy $\varepsilon_{n k}$ and an arbitrary $k$-dependent quantity
$X_{n k}$, and displays the corresponding Fermi surface color coded to indicate the 
value of the given quantity on the Fermi surface.
Furthermore, FermiSurfer can display stereograms, nodal lines of the given quantity,
cross section of the Fermi surface
at any plane, and illuminate either the occupied or unoccupied side of the Fermi surface.
Various first-principles software packages can generate data files readable by FermiSurfer,
several examples of which are used 
to demonstrate the benefit of using FermiSurfer. 

This work was supported by 
Building of Consortia for the Development of Human Resources
in Science and Technology and
Priority Issue (creation of new functional devices and high-performance materials
to support next-generation industries)
to be tackled by using Post `K' Computer
from the MEXT of Japan.
A part of the numerical calculations in this paper were done on the
supercomputer in ISSP at the University of Tokyo.
The authors would like to thank Enago (www.enago.jp) for the English language review.





\bibliographystyle{elsarticle-num}
\bibliography{fermi_ref}







\end{document}